\documentstyle[epsfig]{disk}
\newcommand{\gtrsim}{\mathrel{\hbox{\rlap{\lower.55ex \hbox {$\sim$}}
                   \kern-.3em \raise.4ex \hbox{$>$}}}}
\newcommand{\lesssim}{\mathrel{\hbox{\rlap{\lower.55ex \hbox {$\sim$}}
                   \kern-.3em \raise.4ex \hbox{$<$}}}}

\begin{document}

\title{%
Outbursts of WZ\,Sge stars/TOADs: a phenomenological comparison with 
soft X-ray transients}

\author{Erik KUULKERS \\
{\it Space Research Organization Netherlands, 
Sorbonnelaan 2, 3584 CA Utrecht, \&\ 
Astronomical Institute, Utrecht University, P.O.~Box 80000, 3507 TA 
Utrecht, The Netherlands, E.Kuulkers@sron.nl}}

\maketitle

\section*{Abstract}

The outbursts of WZ\,Sge stars (or TOADs),
are compared to those seen in the (soft) X-ray transients. 
Both types of outbursts exhibit strong similarities:
large amplitudes, long recurrence
times, occurrence of superhumps, and of rebrightenings or reflares at the 
end or just after the main outburst. This suggests that the same 
kind of mechanism is at work to produce
these outbursts. I also briefly discuss their differences 
and whether superoutbursts may exist among the very long
($\gtrsim$month) outbursts in U\,Gem stars.

\section{Introduction}

The recognition that outbursts seen in soft X-ray transients (SXTs, where
the compact star is a neutron star or black hole) show 
similarities to those seen in the unstable class of cataclysmic 
variables (CVs, where the compact star is a white dwarf)\footnote{For 
reviews on CVs and SXTs I refer to e.g., 
Warner 1995 and Tanaka \&\ Lewin 1995, respectively.}
started with the discovery of the brightest X-ray transient so far, 
i.e., A0620$-$00 (Elvis et al.\ 1975). 
The detection at other wavelengths followed soon thereafter
(for a review see Kuulkers 1998). Analysis of archival plates
showed another outburst which occurred $\sim$60~years earlier, in 1917 
(Eachus et al.\ 1976). The large optical outburst amplitude ($\sim$8~mag) and 
long recurrence time were found to be comparable to those of recurrent 
novae, including the dwarf novae (DNe) WZ\,Sge, AL\,Com and VY\,Aqr
(e.g., Eachus et al.\ 1976; Kholopov \&\ Efremov 1976). 
Based on the luminosity of A0620$-$00 in X-rays it was suggested that the compact
star in this system was a neutron star or black hole (e.g., Elvis et al.\ 
1975). Ten years later it was dynamically established to be a black hole
(McClintock \&\ Remillard 1986). 

Mainly the availability of all-sky X-ray monitors increased the sample of 
X-ray transients with well covered outburst light curves (e.g., White et al.\ 1984; for
a recent compendium of SXTs light curves I refer to Chen et al.\ 1997). 
It was noticed by van Paradijs \&\ Verbunt (1984), through a comparison of 
a sample of SXTs and DNe, that they had comparable rise times to maximum, 
and that, therefore, the nature of their outbursts may be similar. 
The optical outburst amplitudes of SXTs range from 2 to 9~mag
(e.g., Shahbaz \&\ Kuulkers 1998), a range also seen
in DNe (e.g., Warner 1995). To date, various other characteristics of dwarf novae 
outbursts are also found in  SXTs, 
strengthening their similarity. A nice example is the delay between the 
start of the outburst in the UV and optical in DNe (e.g., Polidan \&\ Holberg 
1987), which has been observed in the form of an X-ray delay with respect to the
optical in SXTs
(e.g., Orosz et al.\ 1997). 

\section{A phenomenological comparison}

\subsection{SU\,Uma stars}

A subgroup of the DNe, the SU\,UMa stars, shows both short ($\sim$days) and 
long ($\sim$weeks) outbursts.
The long outbursts, called superoutbursts, are much less frequent than the
short outbursts (normal outbursts), and are somewhat brighter ($\sim$1~mag). 
During superoutbursts a unique feature develops: a photometric variation
with a period, P$_{\rm sh}$, a few \%\ longer than the orbital period, i.e.,
the superhump. In Fig.~1, I show the observed superhump period excess, 
$\epsilon$, vs.\ the orbital period, P$_{\rm orb}$, for SU\,UMa stars
(see Patterson 1998 and in these proceedings)\footnote{This sample includes 
systems which are permanently in a high accretion state, so-called 
``permanent superhumpers''.}. 

The recurrence
times of normal outbursts are rather erratic. For superoutbursts 
they are more stable; however, they may change from time 
to time (Vogt 1980). More or less stable recurrence 
times have also been found in various SXTs (see Priedhorsky 1986).
For example, one of them, 4U\,1630$-$47, showed outbursts every 
$\sim$600~days. Recently, it was found that its outburst recurrence time had 
changed to $\sim$680~days, very similar to what is seen in 
SU\,UMa stars (Kuulkers et al.\ 1997). 

At one of the extreme ends of the SU\,UMa stars are the stars with long 
outburst recurrence times; for them almost every outburst is a superoutburst.
Such stars are known as WZ\,Sge stars (e.g., Bailey 1979), sometimes referred
to as TOADs (Howell et al.\ 1995).
It is this subgroup which indeed shows the strongest similarities with the SXTs
(Kuulkers et al.\ 1996; 
see also Lasota 1995). Note that `normal' SU\,UMa stars
may also exhibit long intervals of no outbursts during which the source 
is fainter than normal, such as HT\,Cas (e.g., Robertson \&\ Honeycutt 1996). 
On the other 
hand, some WZ\,Sge stars may show periods of more regular outburst behaviour
during which the source is somewhat brighter in quiescence than normal,
such as BZ\,UMa (Jurcevic et al.\ 1994). 

\subsection{Superhumps in soft X-ray transients}

The detection of photometric variations with a period slightly larger than the
orbital period during outbursts of SXTs immediately pointed to 
superhumps (e.g., Charles et al.\ 1991, Callanan \&\ Charles 1991; 
see O'Donoghue \&\ Charles 1996). In Table~1 I show the SXTs for
which both the orbital period (P$_{\rm orb}$) and $q$ are known.
In three cases the outburst data reveal clear evidence for superhumps with a 
period only 1--2\%\ longer than the orbital period. These three SXTs can 
also be found in Fig.~1. The short superhump period excesses are 
very similar to those found in various WZ\,Sge stars (e.g., Kuulkers 
et al.\ 1996, Patterson et al.\ 1996).

\begin{table}[t]
    \caption{Some properties of various SXTs$^a$} 
\vspace{.5pc}
\begin{center}
\begin{tabular}{l|cccc} \hline
\multicolumn{1}{c|}{SXT} & \multicolumn{1}{c}{P$_{\rm orb}$} & 
\multicolumn{1}{c}{$\epsilon$} & \multicolumn{1}{c}{$q$} & 
\multicolumn{1}{c}{M$_1$} \\
~ & (hr) & {\footnotesize (=$({\rm P}_{\rm sh}$$-$${\rm P}_{\rm orb})/{\rm P}_{\rm orb}$)} & (=M$_2$/M$_1$) & (M$_{\odot}$) \\ \hline
GRO\,J0422+32 (Nova Per '93) & 5.09 & 0.0167$\pm$0.0056 & $<$0.083 & 10$\pm$5 \\
A0620$-$00 (Nova Mon '75) & 7.75 & - & 0.067$\pm$0.004 & 10$\pm$5 \\
GS\,2000+25 (Nova Vul '88) & 8.26 & 0.0096$\pm$0.0009 & 0.042$\pm$0.017 & 10$\pm$4 \\
GS\,1124$-$68 (Nova Mus '91) & 10.40 & 0.0099$\pm$0.0028 & 0.125$\pm$0.031 & 6$^{+5}_{-2}$ \\
Cen\,X-4 (Nova Cen '69) & 15.1 & - & 0.20$\pm$0.04 & 1.3$\pm$0.6 \\
GRO\,J1655$-$40 (Nova Sco '94) & 62.9 & - & 0.28$\pm$0.07 & 7$\pm$1 \\
GS\,2023+34 (V404\,Cyg) & 155 & - & 0.059$\pm$0.003 & 12$\pm$2 \\ \hline
\multicolumn{5}{l}{\footnotesize $^a$~Data taken from O'Donoghue \&\ Charles (1996) and Charles (1998).} \\
\end{tabular}
\end{center}
\end{table}

A correlation between superhump period excess, 
$\epsilon$, and mass ratio, $q$, was 
predicted by Whitehurst (1988) and confirmed by Molnar \&\ Kobulnicky (1992).
As shown by Patterson (1998) this relation is roughly 
$\epsilon = 0.23q/(1+0.27q)$, which allows one to estimate $q$ solely from the 
superhump period excess. Note that the observed values of $q$ in 
GRO\,J0422+32 and GS\,2000+25 agree with this relation, but does not
for GS\,1124$-$68. At the right side of Fig.~1a I supply an additional scaling 
for $q$ based on this relation. 
Theoretically it has been argued that superhumps only appear in systems
with mass ratio $q\lesssim 0.33$ (Whitehurst \&\ King 1991,
Whitehurst 1994). It is, therefore, not surprising that given their values 
for $q$ (Table~1), superhumps are seen in SXTs as well 
(see, e.g., also Whitehurst 1994).
As noted by Patterson (1998) the inferred values of $q$ from 
$\epsilon$ range up to $\sim$0.5 for SU\,UMa stars and permanent superhumpers, 
in contradiction with superhump theory.

\begin{figure}[t]
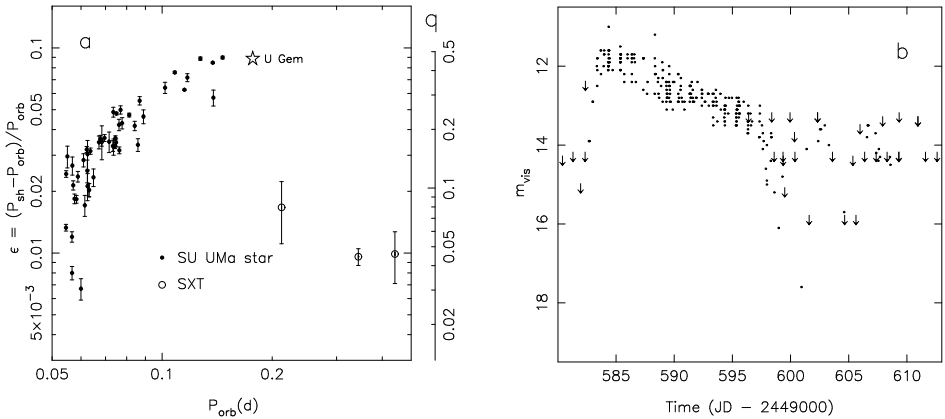

\begin{center}
   \psfig{figure=ek_fig1a.ps,bbllx=52pt,bblly=139pt,bburx=584pt,bbury=626pt,height=5.5cm}
   \psfig{figure=ek_fig1b.ps,bbllx=1pt,bblly=217pt,bburx=562pt,bbury=703pt,height=5.5cm}
\end{center}
   \caption{({\bf a}) The observed superhump period excess, $\epsilon$,
as a function of the orbital period, P$_{\rm orb}$, for SU\,UMa stars 
and permanent superhumpers. Superhump theory predicts a relation between 
$\epsilon$ and the mass ratio, $q$, which is indicated at the right
(after Patterson 1998). Included are the SXTs in which clear evidence for 
superhumps is seen (Table 1). 
The known values of $q$ and P$_{\rm orb}$ for U\,Gem are also plotted; see text.
({\bf b}) The 1994 superoutburst light curve of UZ\,Boo as compiled from the 
AAVSO, BAAVSS, VSNET and VSOLJ reports. Errors denote upper limits. 
This is an update from Kuulkers et al.\ (1996).} 
\end{figure}

\subsection{Superhump growth time and secondary maximum}

Superhumps in WZ\,Sge stars and SXTs have long growth times, i.e., they appear 
one or two weeks, respectively, months, after the outburst has started. 
In `normal' SU\,UMa stars they apppear after one to several days
(e.g., Vogt 1993). This is due to the fact that
the strength of the tidal instability decreases below $q\sim 0.1$
(e.g., Molnar \&\ Kobulnicky 1992, Whitehurst 1994). 

The start of the superhumps in at least GRO\,J0422+32 was seen to be 
very close to the start of the secondary maximum, which suggests
a possible connection between the two (Kato et al.\ 1995).
The same can be inferred for some of the WZ\,Sge stars, e.g., 
AL\,Com and EG\,Cnc (see Patterson et al.\ 1996, 1998).
It was, therefore, suggested that the origin of the superhumps and
secondary maximum are the same for both the SXTs and WZ\,Sge stars
(Kuulkers et al.\ 1996).

\subsection{Rebrightenings/reflares}

Reflares after the main outburst or sudden drops with a rebrightening
near the end of the outburst are present in several 
WZ\,Sge stars (see Richter 1992, Kuulkers et al.\ 1996; e.g., UZ\,Boo: Fig.~1b).
The best example of reflares is EG\,Cnc which showed
up to six `normal' outbursts after its main outburst (Patterson et
al.\ 1998, Kato et al.\ 1998). 
The best example of a rebrightening is WZ\,Sge itself,
which showed a fast drop near the end of the 1978 outburst but came back to 
stay on for a longer time. It then gradually decayed to quiescence
(Ortolani et al.\ 1980, Patterson et al.\ 1981). Reflares/rebrightenings are,
however, not exclusively seen in WZ\,Sge stars: 
SU\,UMa also showed a reflare after its 1989 outburst (Udalski 1990).

In SXTs reflares have been observed as well. GRO\,J0422+32 is the best known
example (Callanan et al.\ 1995, Chevalier \&\ Ilovaisky 1995): it showed two 
strong and various weaker outbursts after its main outburst.
The SXT 4U\,1543$-$47 showed multiple
rebrightenings/reflares in X-rays during its 1971/1972 outburst 
(Li et al.\ 1976). In GRO\,J0422+32 the two strong reflares
apeared $\sim$120~days after each other, which is {\em similar} to the time
from maximum to the secondary maximum of the outburst (e.g., 
Callanan et al.\ 1995). This also holds for the reflares seen in the 
WZ\,Sge stars (see Kuulkers et al.\ 1996).
A good example is again EG\,Cnc (see Kuulkers \&\ Howell 1996): 
a clear secondary maximum can be seen 
about a week after outburst maximum, whereas the reflares appeared 
a week apart (see, e.g., Patterson et al.\ 1998). 
This may indicate a certain clock which operates in both the SXT and WZ\,Sge 
stars which controls the timing of the secondary maximum (and/or onset
of superhumps) and the time between reflares.

\section{Discussion}

\subsection{Differences}

Although I have shown here that there are similarities in the outburst 
behaviour of SXTs and WZ\,Sge stars/TOADs, obvious differences are present
as well.
The main difference is the fact that SXTs harbour a neutron star or a black 
hole instead of a white dwarf. A black hole ($\gtrsim$3\,M$_{\odot}$) can
easily account for the observed low values of $q$. 
The presence of such compact stars leads to more energetic outbursts, 
especially in the X-ray and
$\gamma$-ray region. Strong radiation at these wavelengths almost certainly
irradiate the disk and/or secondary star and may influence the duration and/or
brightness of the outbursts of SXTs (e.g., van Paradijs \&\ Verbunt 1984,
van Paradijs \&\ McClintock 1994, Shahbaz \&\ Kuulkers 1998).
In DNe, irradiation of the accretion disk is probably not very important
(e.g., Hameury et al.\ 1998). However, the hot white dwarf, heated during
outburst, may influence the very cool (e.g., Ciardi et al.\ 1998) 
secondary star in WZ\,Sge stars.
This could account for the presence of reflares/rebrightenings in 
such systems (Warner 1998, see also Lasota in these proceedings).
In this respect I note the striking similarity in the outburst amplitude
and morphology of the outburst light curves of A0620$-$00 and AL\,Com
(Kuulkers 1998). The only difference is the timescale of their outburst 
duration by a factor $\sim$5. If irradiation influences 
the outburst brightness in A0620$-$00 it remains unclear why a similar
outburst amplitude can be seen in AL\,Com; maybe irradiation has less effect
on brightness than expected.

\subsection{Superoutbursts in U\,Gem stars?}

\begin{figure}[t]
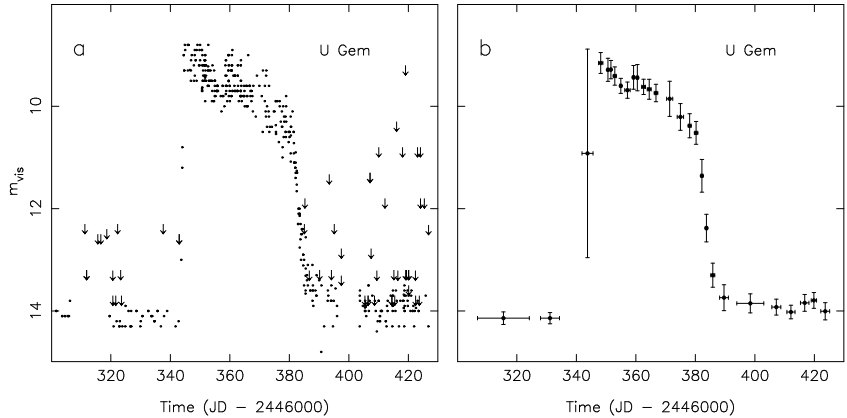

\begin{center}
   \psfig{figure=ek_fig2a.ps,bbllx=58pt,bblly=217pt,bburx=562pt,bbury=700pt,clip=yes,height=5.5cm}
   \psfig{figure=ek_fig2b.ps,bbllx=100pt,bblly=217pt,bburx=562pt,bbury=700pt,clip=yes,height=5.5cm}
\end{center}
   \caption{({\bf a}) Visual estimates of the 1995 outburst of U\,Gem done
by the AAVSO, AFOEV, BAAVSS and VSOLJ. Arrows denote upper limits. ({\bf b})
Averages (15 consecutive points) of the outburst light curve of ({\bf a}). The errors 
are the mean absolute deviations per bin (see Press et al.\ 1992).}
\end{figure}

Very long ($\gtrsim$month) outbursts seem to be very rare among DNe above the 
period gap, where most of them belong to the U\,Gem class.
Lasota (1995) made me aware of a `sketch' in Mason et al.\ (1988) of a long 
($\sim$40~days)
outburst of U\,Gem which occurred in 1985. This was not seen
before in this system; the outburst durations are normally bimodally distributed
($\sim$9 and $\sim$17~days; see, e.g., van Paradijs 1983). 
Fig.~2a shows a compilation of the available visual estimates 
during that time. The long outburst is reminiscent to that 
seen in WZ\,Sge stars. The individual measurements suggest a 
secondary maximum after more than a week, as can be seen in some WZ\,Sge star 
outburst light curves (Sect.~2.4). 
However, statistically this may not be significant 
(Fig.~2b). It is interesting to speculate that this long outburst is in a
fact a superoutburst and thus may have exhibited superhumps.
A crazy idea is to put U\,Gem in Fig.~1a, given
$q=0.46\pm 0.03$ and P$_{\rm orb}=4.25$\,hr (Friend et al.\ 1990).
Surprisingly, the inferred value of $\epsilon$ of U\,Gem
($\sim$9.5\%\/) is close to the extreme end of the $\epsilon$ 
vs.\ P$_{\rm orb}$ relation. 
No observations have been done 
during the 1985 outburst of U\,Gem to confirm the presence 
of superhumps.
It has been argued earlier that the long outbursts in U\,Gem stars may in 
fact be superoutbursts (van Paradijs 1983). However, no superhumps have been seen 
in the long outbursts ($\sim$2 weeks) of e.g., SS\,Cyg, despite 
considerable coverage (Honey et al.\ 1989).
Moreover, the recurrence time properties of the long outbursts in U\,Gem 
stars are different to that of the superoutbursts in SU\,UMa stars (see 
Warner 1995). It is interesting to note (see Warner 1995)
that the only SU\,Uma star so far 
above the period gap, i.e., TU\,Men (P$_{\rm orb}=2.82$\,hr), has a trimodal
distribution of outburst widths, i.e., short ($\sim$1~day), long ($\sim$8~days)
and superoutbursts ($\sim$20~days). It could therefore be that U\,Gem's
outburst duration distribution also follows a trimodal distribution, where
superoutbursts are very rare.
  
Among the WZ\,Sge star candidates a few
have (inferred) orbital periods above and just below the period gap
(see, e.g., O'Donoghue et al.\ 1991, Howell et al.\ 1995).
These may be candidates of systems which have evolved way past the 
period minimum (Howell et al.\ 1997; see, however, Patterson 1998). 
Among these are EF\,Peg (Howell et al.\ 1995); superhumps were seen during its 
1991 outburst with P$_{\rm sh}\simeq 2.05$\,hr (Howell \&\ Fried 1991, Kato \&\ Takata 1991). 
The candidates above the period gap were e.g.\ DO\,Dra, AR\,And and WW\,Cet
(Richter 1985, O'Donoghue et al.\ 1991), Howell et al.\ 1995). 
They seemed to have long outburst recurrence times
and large amplitude outbursts. However, a closer look
of the candidates above the period gap have revealed that they are normal U\,Gem stars 
(see Ritter \&\ Kolb 1998, and references therein).
Future observations are needed to see whether EF\,Peg is a real WZ\,Sge star,
and whether WZ\,Sge stars indeed exist above the period gap. Such observations
have large implications on our understanding of evolution of CVs among the 
period minimum (Howell et al.\ 1997, Patterson 1998).
Moreover, such stars, 
having larger orbital periods, might then be even closer cousins to the 
SXTs.

~\\
{\bf Acknowledgements}:
In this paper we have used, and acknowledge with thanks, data from the AAVSO 
International Database, based on observations submitted to the AAVSO by 
variable star observers worldwide, as well as observations of
the AFOEV, BAAVSS and VSOLJ, and from VSNET.
Most of this work is entirely attributed to ``the zeal and assiduity, not only 
of the public observers, but also of the many private individuals, who nobly
sacrifice a great portion of their time and fortune to the laudable pursuit"
of these variables (Moyes 1831).

\section{References}


\footnotesize

\begin{center}
\begin{tabular}{ll}
Bailey J. 1979, MNRAS 189, 41P & O'Donoghue D., et al. 1991, MNRAS 250, \\
Callanan P.J., Charles P.A. 1991, MNRAS & $^{~~~}$363 \\
$^{~~~}$249, 573 & Orosz J., et al. 1997, ApJ 478, L830 \\
Callanan P.J., et al. 1995, ApJ 441, 786 & Ortolani S., et al. 1980, A\&A 87, 31 \\
Charles P.A. 1998, in: Theory of Black & Patterson J. 1998, PASP 110, 1132 \\
$^{~~~}$Hole Accretion Disks, CUP, p.~1 & Patterson J., et al. 1981, ApJ 248, 1067 \\
$^{~~~}$(astro-ph/9806217) & Patterson J., et al. 1996, PASP 108, 748 \\
Charles P.A., et al. 1991, MNRAS 249, 567 & Patterson J., et al. 1998, PASP 110, in press \\
Chen W., et al. 1997, ApJ 491, 312 & Polidan R.S., Holberg J.B. 1987, MNRAS 225, 131 \\
Chevalier C., Ilovaisky S.A. 1995, A\&A 297, 103 & Press W.H., et al. 1992, in: Numerical Recipes  \\
Ciardi D.R., et al. 1998, ApJ 504, 450 & $^{~~~}$2nd ed., CUP \\
Eachus L.J., et al. 1976, ApJ 203, L17 & Priedhorsky W. 1986, A\&SS 126, 89 \\
Elvis M., et al. 1975, Nat 257, 656 & Richter G.A. 1985, AN 307, 221 \\
Friend M.T., et al. 1990, MNRAS 246, 637 & Richter G.A. 1992, in: Vi\~na Del Mar Workshop  \\
Hameury J.-M., et al. 1998, MNRAS, in & $^{~~~}$on Cataclysmic Variable Stars, ASP Conf. \\
$^{~~~}$press (astro-ph/9808347) & $^{~~~}$Proc.\ 29, p.~12 \\
Honey W.B., et al. 1989, MNRAS 236, 72 & Ritter H., Kolb U. 1998, A\&AS 129, 83 \\
Howell S.B., Fried R. 1991, IAUC 5372 & Robertson J.W., Honeycutt R.K. 1996, AJ \\
Howell S.B., et al. 1995, ApJ 439, 337 & $^{~~~}$112, 2248 \\
Howell S.B., et al. 1997, MNRAS 287, 929 & Shahbaz T., Kuulkers E. 1998, MNRAS 295, L1 \\
Jurcevic J.S., et al. 1994, PASP 106, 481 & Tanaka Y., Lewin W.H.G. 1995, in: X-ray \\
Honey W.B., et al. 1989, MNRAS 236, 727 & $^{~~~}$Binaries, CUP, p.~126 \\
Kato T., Takata T. 1991, IAUC 5371 & Udalski A. 1990, AJ 100, 226 \\
Kato T., et al. 1995, PASJ 47, 163 & van Paradijs J. 1983, A\&A 125, L16 \\
Kato T., et al. 1998, PASJ, in press & van Paradijs J., McClintock J.E. 1994, A\&A \\
Kholopov P.N., Efremov Yu.N. 1976, PZ 20,  & $^{~~~}$290, 133 \\
$^{~~~}$277 & van Paradijs J., Verbunt F. 1984, in: High \\
Kuulkers E. 1998, NewAR 42, 1 (astro-ph/ & $^{~~~}$Energy Transients, AIP Conf.\ Proc.\ 115, p.~49 \\
$^{~~~}$9805031) & Vogt N. 1980, A\&A 88, 66 \\
Kuulkers E., Howell S.B. 1996, VSNET-alert & Vogt N. 1993, in: Cataclysmic Variables \\
$^{~~~}$654 & $^{~~~}$and Related Objects, Ann.\ Israel Phys.\ Soc.\\
Kuulkers E., et al. 1996, ApJ 462, L87 & $^{~~~}$10, p.~63 \\
Kuulkers E., et al. 1997, MNRAS 291, 81 & Warner B. 1995, in: Cataclysmic Variable Stars, \\
Lasota J.-P. 1995, in: Compact Stars in & $^{~~~}$CUP \\
$^{~~~}$Binaries, IAUS 165, p.~43 & Warner B. 1998, in: Wild Stars in the Old \\
Li F.K., et al. 1976, ApJ 203, 187 & $^{~~~}$West, ASP Conf.\ Ser.\ 137, p.~2 \\
Mason K.O., et al. 1988, MNRAS 232, 779 & White N.E., et al., 1984, in: High Energy \\
McClintock J.E., Remillard R.A. 1986, ApJ & $^{~~~}$Transients, AIP Conf.\ Proc.\ 115, p.~31 \\
$^{~~~}$308, 110 & Whitehurst R. 1988, MNRAS 232, 35 \\
Molnar L.A., Kobulnicky H.A. 1992, ApJ 392, 678 & Whitehurst R. 1994, MNRAS 266, 35 \\
Moyes J. 1831 (Feb.\ 9, 1827), MNRAS 1, 1 & Whitehurst R., King A.R. 1991, MNRAS 249, \\
O'Donoghue D., Charles P.A. 1996, MNRAS & $^{~~~}$25 \\
$^{~~~}$282, 191 & \\
\end{tabular}
\end{center}

\end{document}